\documentclass{ws-ijmpb}
\usepackage{hyperref}
\usepackage{subfigure}
\usepackage{graphicx}

\usepackage{xy}
\xyoption{matrix}
\xyoption{frame}
\xyoption{arrow}
\xyoption{arc}

\usepackage{ifpdf}
\ifpdf
\else
\PackageWarningNoLine{Qcircuit}{Qcircuit is loading in Postscript mode.  The Xy-pic options ps and dvips will be loaded.  If you wish to use other Postscript drivers for Xy-pic, you must modify the code in Qcircuit.tex}
\xyoption{ps}
\xyoption{dvips}
\fi

\entrymodifiers={!C\entrybox}

\newcommand{\bra}[1]{{\left\langle{#1}\right\vert}}
\newcommand{\ket}[1]{{\left\vert{#1}\right\rangle}}
\newcommand{\qw}[1][-1]{\ar @{-} [0,#1]}
\newcommand{\qwx}[1][-1]{\ar @{-} [#1,0]}

\newcommand{\gate}[1]{*+<.6em>{#1} \POS ="i","i"+UR;"i"+UL **\dir{-};"i"+DL **\dir{-};"i"+DR **\dir{-};"i"+UR **\dir{-},"i" \qw}
\newcommand{\meter}{*=<1.8em,1.4em>{\xy ="j","j"-<.778em,.322em>;{"j"+<.778em,-.322em> \ellipse ur,_{}},"j"-<0em,.4em>;p+<.5em,.9em> **\dir{-},"j"+<2.2em,2.2em>*{},"j"-<2.2em,2.2em>*{} \endxy} \POS ="i","i"+UR;"i"+UL **\dir{-};"i"+DL **\dir{-};"i"+DR **\dir{-};"i"+UR **\dir{-},"i" \qw}

\newcommand{\control}{*!<0em,.025em>-=-<.2em>{\bullet}}

\newcommand{\ctrl}[1]{\control \qwx[#1] \qw}

\newcommand{\multigate}[2]{*+<1em,.9em>{\hphantom{#2}} \POS [0,0]="i",[0,0].[#1,0]="e",!C *{#2},"e"+UR;"e"+UL **\dir{-};"e"+DL **\dir{-};"e"+DR **\dir{-};"e"+UR **\dir{-},"i" \qw}
\newcommand{\ghost}[1]{*+<1em,.9em>{\hphantom{#1}} \qw}

\newcommand{\lstick}[1]{*!R!<.5em,0em>=<0em>{#1}}

\newcommand{\Qcircuit}{\xymatrix @*=<0em>}

\begin{document}

\markboth{DAVIDE GIROLAMI, RUGGERO VASILE, and GERARDO ADESSO}
{THEORETICAL INSIGHTS ON MEASURING QUANTUM CORRELATIONS}

%
\catchline{}{}{}{}{}
%

\title{THEORETICAL INSIGHTS ON MEASURING QUANTUM CORRELATIONS}

\author{DAVIDE GIROLAMI $^{\dag\ddag,}$, RUGGERO VASILE$^{\dag\S}$, and GERARDO ADESSO$^{\dag,\P}$}

\address{$^{\dag}$School of Mathematical Sciences, The University of Nottingham, University Park, Nottingham NG7 2RD, United Kingdom\\
$^\ddag$Corresponding author. \texttt{pmxdg1@nottingham.ac.uk} \\
$^{\S}$ \texttt{ruggero.vasile@nottingham.ac.uk}\\
$^{\P}$ \texttt{gerardo.adesso@nottingham.ac.uk}}

\maketitle

\begin{history}
\received{25 May 2012}
\end{history}

\begin{abstract}
We review a recently developed theoretical approach to the experimental detection and quantification of bipartite quantum correlations between a qubit and a $d$ dimensional system. Specifically, introducing a properly designed measure $Q$, the presented scheme allows us to quantify general quantum correlations for arbitrary states of  $2\otimes d$ systems without the need to fully reconstruct them by tomographic techniques. We take in exam the specifics of the required experimental architecture in nuclear magnetic resonance and optical settings. Finally we extend this approach to models of open system dynamics and discuss possible advantages and limitations in such a context.
\end{abstract}


\section{Introduction}
In recent years quantum information has been identified as an ideal ground to test the robustness of the fundamental principles of quantum mechanics and, at the same time, to push technology over its inherent limits.
At the core of this new research field, a fundamental role is played by one of the key features of the quantum world, the superposition principle, which allows for the existence of a class of physical states without any classical analogue. Given a composite bipartite system, one can engineer quantum states of the global system in which the subsystems are not in a well defined state and share some amount of genuinely quantum correlations, e.g. entanglement. Quantum entanglement is recognized as an essential ingredient for most of quantum information protocols, and, therefore, a fundamental resource for quantum technology \cite {ent}.\par
However, the last decade has seen a great deal of attention pointed to a subtler but more general kind of quantum correlations (QCs from now on). For instance, it has been established that separable (i.e., unentangled) mixed states can still possess correlations of a quantum nature, which cannot be described within a classical probabilistic framework \cite{OZ,HV,perinotti}.  It has been conjectured that such general QCs play a role in performing better-than-classical algorithms for specific computational tasks \cite{laf1,dattabarbieriaustinchaves,white,chaves,genio,dattamerging,streltsov}, and they have been identified as a figure of merit for remote state preparation \cite{chaves,rspnature,rspibrido} and phase estimation protocols \cite{modix}. An intense experimental activity followed from these theoretical findings \cite{white,laf2,mazi,cinesiwitness,cinesinatcomm,ginanuovo}. More generally, reaching a full understanding of the deep nature of the QCs, finding out how we can exploit them, testing their accessibility and resilience under decoherence, and discovering which role they play in the study of complex and many body systems, are among the most exciting scientific challenges linking quantum information theory, complexity theory and condensed matter physics \cite{nature}.  \par
Qualitatively, general QCs are related to the disturbance induced by the measurement process on a physical system \cite{MID,luofu}, while concepts and tools employed from information theory allow a quantitative evaluation of the amount of QCs in the state of the system. Several QCs measures have been introduced \cite{OZ,HV,dakic,altremisure} and appealing operational interpretations have been associated to them \cite{revmodi}. At this stage, the natural next step should be to establish a link between theoretical and experimental quantification of QCs, but, unfortunately, all the QCs measures are defined by means of a state-dependent optimization and are not directly associable with observable quantities, i.e. Hermitian operators. Since full state reconstruction is a tedious, if not unfeasible, procedure, it becomes desirable to find ways to evaluate QCs by means of a smaller number of measurements than the ones required by tomographic techniques. In this direction, appreciable attempts to detect nonvanishing QCs by observable witnesses  have been realized \cite{laf2,mazi,cinesinatcomm}. Anyway, reminding that almost all states possess QCs \cite{acinferraro} and that \emph{the} pivotal question around QCs is whether they are exploitable as resource, it seems worthy to pursue a more informative (but still experimentally manageable) quantitative characterization of QCs. \par
In this article, we review and update the proposal two of us presented in Ref.~\refcite{prl}, which bridges the gap between theoretical and experimental evaluation of QCs. First, QCs quantifier $Q$ for two-qubit states is introduced, directly derived from the geometric discord defined in Ref.~\refcite{dakic}, which is given by a state-independent expression not involving any optimization procedure. Also, a generalization of $Q$ to detect bipartite QCs for states of $2\otimes d$ dimensional systems is provided. In particular, as example given, we focus on the amount of QCs produced in the four-qubit realization of the DQC1 protocol \cite{laf1,laf2,ginanuovo}, which recently captured the interest of the experimental community \cite{laf2,ginanuovo,altremisure}. Noting that there is no need for a complete reconstruction of the state in order to calculate $Q$,  we explore the possible direct implementation of the non-tomographic measurement scheme required to detect the value of $Q$ on an unknown state. In particular, we express $Q$ as a function of observable quantities $\langle O_i\rangle$, i.e. the expectation values of proper Hermitian operators $O_i$. The nature and the number of such operators is obviously dependent on the particular setting considered. \par Two possibilities are taken in exam: first, we consider the NMR (Nuclear Magnetic Resonance) setting \cite{brazirev}, in which we obtain information on the system by means of spin measurements. Here, our protocol allows a gain (over full state tomography) which is linearly dependent on Bob's dimension. Then, we focus on the quantum optical setup, which implies to recast $Q$ in terms of expectation values of projectors and swap operators. The number of measurements required is independent of Bob's dimension $d$, while the complexity of the setting, i.e., the number of optical devices needed, increases only linearly with $d$. Thus, an exponential gain over tomography can be achieved in terms of required resources. The quantum circuits simulating the measurements can be designed following a well established literature \cite{paisa12,paz,brun,ekert,filip1,filip2,winter,horo0}. In spite of a minor advantage in terms of number of measurements, the NMR implementation is by far easier to realize than the optical one \cite{laf2,mazi,ginanuovo}.\par
The advantage provided by the correlation quantifier $Q$ is even more striking in dynamical contexts. To experimentally study the evolution of QCs, many sets of measurements need to be performed at different instants of time, thus the number of measurements required to resolve the dynamics increases dramatically. The introduction of $Q$ and of its measurement schemes then helps in reducing the number of necessary resources compared, for instance, to those needed for full dynamical tomographic reconstruction. In view of this  possibility we provide a brief discussion on the properties of the quantum correlation quantifier $Q$ in the context of open quantum systems \cite{BrePet,Weiss}.  Due to the prominent role that QCs promise to play for realistic quantum technology, in recent years strong efforts have been made to provide a qualitative and quantitative picture on their evolution under various types of decoherence. The reader can find some relevant references of interest in the case of entanglement \cite{YuEb,Prau,ManParOli,Paz,mazzola3,Vasile1,Vasile2,LoFraDisc}, quantum discord \cite{Vasile2,LoFraDisc,DiscQubit,Maziero,fanchini,PazDisc}, and other quantifiers of quantum correlations dynamics \cite{GeoOpen,ReDiOp}.

Here we will focus on two different system-reservoir models for two-qubits systems. First we consider a scenario with tunable non-Markovianity where each qubit is coupled with a local Lorentzian environment \cite{BelLoCom}, and we find that the quantifier $Q$ witnesses typical non-Markovian dynamical features for quantum correlation, e.g. presence of oscillations due to the non-divisibility of the associated quantum dynamical map \cite{breuerthe,plenio}. The second example is instead a Markovian scenario where each qubit decays accordingly to a non-dissipative dynamics \cite{LauraPRL}. Recently this system has been invoked as a paradigmatic example showing a transition between a quantum and a classical decoherence regime, i.e. the presence of a time interval where quantum correlations quantified by the entropic quantum discord \cite{OZ,HV} do not decay. Other quantifiers of correlations like the geometric discord \cite{dakic} also witness a change in the decay properties of quantum correlations, despite showing a different qualitative behavior. If the quantifier $Q$ is instead employed, however, such transition is not evident, due to the very definition of the quantity $Q$ which does not involve minimization procedures. Anyway, phenomena such as frozen discord are typical of states identified with a small number of parameters, e.g., Werner states, evolving under purpose-driven dynamics preserving such peculiarity. Obviously, the state reconstruction in such occurrences does not represent a serious concern, if a prior knowledge on the form of the states under consideration is assumed. Therefore the advantages of adopting $Q$ rather than more complex measures of QCs cannot be apparent in those settings.

The paper is organized as follows. In Section~\ref{two}, after a brief review on geometric discord $D_G$, we present the definition of the experimentally accessible bipartite QCs quantifier  $Q$. We initially consider the paradigmatic two-qubit case, then a straightforward extension to $2\otimes d$ dimensional systems is provided, and an explicit analysis carried out for the four-qubit DQC1 model \cite{laf1}. Section~\ref{three} describes possible implementations of the experimental schemes required to evaluate $Q$ in  laboratory. In particular, features of NMR and optical settings  are explored. In Section~\ref{four} we focus on the open system dynamics of the QCs quantifiers $Q$ and $D_G$ in the two-qubit case, comparing their evolutions for local Markovian and non-Markovian channels.
Finally, we summarize in Section~\ref{con} the main points of our work, with an overview of future developments in the field.

\section{Observable measure of QCs}\label{two}

\subsection{Geometric Discord}
Hereafter we focus on states $\rho$ of a bipartite system $AB$. A quantitative assessment of QCs is inherently dependent on the specific measurement we are going to make on the system. According to a conventional choice stated by the literature of the field, we consider a local measurement on one of the subsystems, say Alice.  The states left undisturbed by such a measurement are called "classical-quantum" states \cite{acinferraro,piani}, and form a null-measure subset $\Omega$ in the set of all density operators. Their density matrix takes the following form
\begin{equation}\label{cq}
\rho_{CQ} = \sum _i  p_i  |i\rangle  \langle i |\otimes\rho _{Bi} ,
\end{equation}
where the positive coefficients $p_i$ define a probability distribution, $\rho_{Bi}$ are density operators of the subsystem $B$ and  $\{|i\rangle\}$ is an orthonormal basis for the Hilbert space of subsystem $A$. One can intuitively define the amount of QCs of a generic state in terms of its minimal distance from the set $\Omega$ of classical-quantum states. Indeed, the geometric discord $D_G$, introduced for the two-qubit case in Ref.~\refcite{dakic}, is defined as
\begin{equation}\label{dgeom}
  D_G(\rho)=2\min_{\rho_{CQ} \in \Omega}  \|\rho -\rho_{CQ}\|_2^2\,,
  \end{equation}
where the Hilbert-Schmidt norm is employed as measure of distance between states, i.e. $\|M\|_2=\sqrt{\text{Tr}(M M^\dagger)} = \sqrt{\sum_i m_i^2}$, and $\{m_i\}$ are the eigenvalues of the matrix $M$. It should also be remarked that a normalization factor $2$ is added in the definition \eqref{dgeom} , in order to obtain $1$ as the maximum value  for the $D_G$, i.e., in the case of Bell states. \par

Geometric discord enjoys two nice theoretical interpretations. First, it quantifies the disturbance induced by local Von Neumann projective measurements  $\Pi=\Pi_A\otimes \mathbb{I}_B$ on the subsystem $A$ \cite{luofu}
 \begin{equation}
  D_G(\rho)=2 \min_{\Pi}  \|\rho -\Pi(\rho) \|_2^2\,.
  \end{equation}
Moreover, it can be recast as the distance of a state from itself after the action of a `root-of-unity' local unitary operation on Alice  $U=U_A\otimes \mathbb{I}_B$  \cite{gharibian}
  \begin{equation}\label{sev}
  D_G(\rho)=2 \min_{U}  \|\rho -U\rho U^{\dagger} \|_2^2\,.
  \end{equation}
Recently $D_G$ has been found out a reliable figure of merit for remote state preparation \cite{rspnature,rspibrido}.  It is appropriate to remind that $D_G$ and all the discord-like measures are not symmetric under subsystems swapping, i.e., performing the measurement on Bob rather than on Alice would lead to define another  class of QCs signatures.\par
Geometric discord can be easily calculated for two qubits. First, one has to write the state in the Bloch-Fano picture\cite{bloch,fano}
 \begin{eqnarray}
   \rho &=& \frac14 \sum_{i,j=0}^3 R_{ij} \sigma_i \otimes \sigma_j\nonumber\\
    &=& \frac 14\left(\mathbb{I}_{4}+\sum_{i=1}^3 x_i\sigma_i \otimes \mathbb{I}_{2} +\sum_{j=1}^3 y_j \mathbb{I}_{2}\otimes \sigma_j+\sum_{i,j=1}^3 t_{ij} \sigma _i\otimes\sigma_j\right),
   \end{eqnarray}
where $R_{ij}=\text{Tr}[\rho(\sigma_i\otimes \sigma_j)]$, $\sigma_0=\mathbb{I}_{2}$, $\sigma _i$ ($i=1,2,3$) are the Pauli matrices, $\vec{x}=\{x_i\},\vec{y}=\{y_i\}$ are the Bloch column vectors associated to $A,B$, and  $t_{ij}$ are the entries of the correlation matrix $t$. Then one has
\begin{eqnarray} \label{eq1}
 D_G(\rho)&=& \frac 12(\|\vec x\|^2 + \|t\|_2^2 -4 k_{\max})\nonumber\\
 &=&2 (\text{Tr}[S]-k_{\max}),
\end{eqnarray}
  where $k_{\max}$ is the largest eigenvalue of the matrix and $S = \frac 14(X+T)$, with $X=\vec x {\vec  x}^T, T=  t t^T$.  From now on we will imply $D_G\equiv D_G(\rho)$, assuming the same convention for all the other quantities.

\subsection{Observable lower bound}
The minimization remaining in the definition of $D_G$ can be actually solved in closed form. Since the characteristic equation of the matrix $S$ is a cubic with real coefficients and roots, is easily solvable by following standard techniques \cite{cubica}. Indeed, the eigenvalues of $S$ can be found by solving an equation of the form
\begin{eqnarray}
k^3+a_0 k^2+a_1 k+a_2=0,
\end{eqnarray}
where
\begin{eqnarray}
a_0&=&-\text{Tr}[S]\nonumber\\
a_1&=&\frac12(\text{Tr}[S]^2-\text{Tr}[S^2])\nonumber\\
a_2&=&-\frac13 ( a_1\text{Tr}[S]+a_0\text{Tr}[S^2]+\text{Tr}[S^3]).
\end{eqnarray}
Introducing the following variables
\begin{eqnarray}
q&=&\frac19 (3a_1-a_0^2)\nonumber\\
r&=&\frac1{54}(9 a_0 a_1-27 a_2-2 a_0^3)\nonumber\\
\theta &=&\arccos\left[\frac r{\sqrt{-q^3}}\right],
\end{eqnarray}
 after a bit of algebra, one obtains
\begin{eqnarray}\label{10}
k_i&=&\frac13\left(\text{Tr}[S]+2 \sqrt{-q^3}\cos{\left[\frac{\theta+\alpha_i}{3}\right]}\right)\nonumber\\
q&=& -\sqrt[3]{\frac 14 (6\text{Tr}[S^2]-2\text{Tr}[S]^2)} \nonumber\\
\theta &=& \arccos\left[(2 \text{Tr}[S]^3-9 \text{Tr}[S]\text{Tr}[S^2]+9\text{Tr}[S^3])
\sqrt{2/(3\text{Tr}[S^2]-\text{Tr}[S]^2)^3}\right]\nonumber\\
\{\alpha_i\}&=&\{0, 2\pi,4\pi\}.
\end{eqnarray}
We now have state independent expressions for the eigenvalues of $S$. Also, we observe that $\theta$ is an arccosine function, thus its domain is  $0 \leq  \theta/3\leq \pi /3$.  The maximum of $\cos\left[\frac{\theta+\alpha_i}{3}\right]$ is therefore always reached for $\alpha_i\equiv\alpha_1=0$. Hence, $k_{\max}\equiv\text{max}\{k_i\}= k_1$, and the geometric discord for an arbitrary two-qubit state $\rho$ can be recast as an explicit function of the coefficients $(\rho_{ij})$
\begin{eqnarray}\label{dg}
D_G &=&2 (\text{Tr}[S]-k_1)\nonumber\\
&=&\frac23\left(2\text{Tr}[S]-\sqrt{6\text{Tr}[S^2]-2\text{Tr}[S]^2}\cos\left[\frac\theta3\right]\right).
\end{eqnarray}
At this point, we remind that the aim is to find an observable QCs measure, i.e. to quantify correlations in terms of observable quantities. The geometric discord in expression Eq.~(\ref{dg}) is just a function of polynomials of the density matrix entries. Protocols for writing linear and even non-linear functionals of $\{\rho_{ij}\}$ in terms of expectation values of Hermitian unitary operators have been extensively developed. Furthermore quantum circuits estimating such quantities have been already designed \cite{cina}. For an overview of the state of the art of the field see Ref.~\refcite{paz}--\refcite{horo0}. Thus in principle nothing prevents us from measuring geometric discord in actual experimental setups.  Unfortunately, in practice, the implementation of the  required architecture seems rather challenging, hence it can be valuable to make a further effort and trying to define a QCs measure endowed with an even simpler and more accessible experimental evaluation.\par
Moved by the previous considerations, we observe that in Eq.~(\ref{dg}) one can fix $\theta=0$ and define \cite{prl}
\begin{eqnarray}\label{q}
Q=\frac23\left( 2\text{Tr}[S]-\sqrt{6\text{Tr}[S^2]-2\text{Tr}[S]^2} \right).
\end{eqnarray}
\begin{figure}[bt]
\centerline{\psfig{file=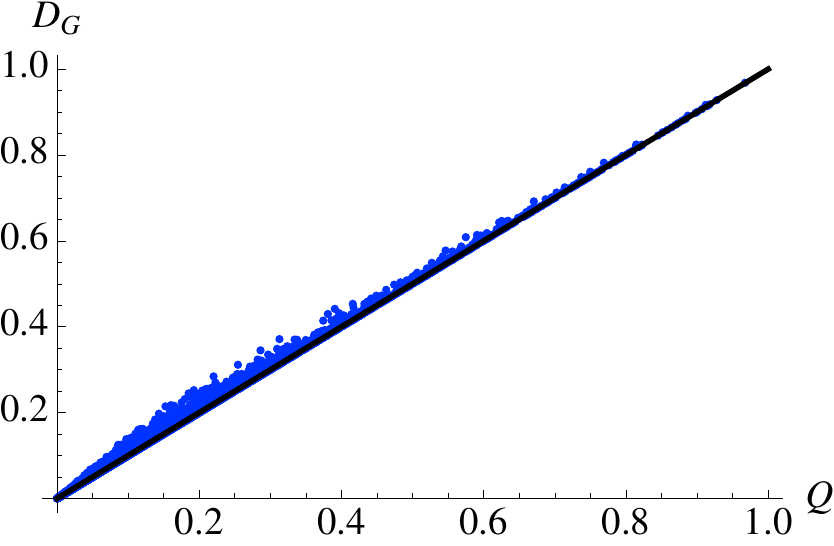,width=8cm}}
\vspace*{8pt}
  \caption{Geometric discord $D_G$ versus $Q$. Sample of $10^4$ randomly generated two-qubit states. The plotted quantities are dimensionless.}
    \label{geo}
\end{figure}
It is immediate to see, from the properties of the cosine, that $Q\le D_G$. In Fig.~\ref{geo} we compare the two quantities, showing that $Q$ is a very tight  lower bound of geometric discord. More important, $Q$ is still a faithful QCs measure. Indeed, the following properties hold.
 \begin{itemize}
 \item $Q\geq 0$, being zero only for classical-quantum states $\rho_{CQ}$, i.e., $Q=0 \iff D_G =0$. To prove this, notice that the condition for vanishing $Q$ is  $\text{Tr}[S]^2=\text{Tr}[ S^2]$. By the Cayley-Hamilton theorem, this implies $\text{Tr}[S]^3=\text{Tr}[S^3]$ and consequently $D_G=0$.
 \item For pure states, $Q$ is equal to geometric discord, as it can be easily proven. The Schmidt decomposition of a pure state $\rho_ p$ of a two-qubit system reads
\begin{eqnarray}
\rho_p=\sum_{ij=0,1} \sqrt{\alpha_i} \sqrt{\alpha_j} |ii\rangle\langle jj|,
\end{eqnarray}
where $\{\alpha_i\}$ are the Schmidt coefficients and $\sum_i \alpha_i=1$. Simple algebraic steps return $ \theta_{p}=0$, thus $D_{G}(\rho_{p})=Q({\rho_{p}})$.
\end{itemize}
In general, for two-qubit states, numerical and partial analytical evidences show that the following chain of inequalities holds: $D_G \geq Q \geq {\cal N}^2$, where $\cal{N}$ is the negativity, a computable entanglement monotone \cite{ent}. All those quantities coincide on pure states, completing the picture presented in Ref.~\refcite{interplay}.  For a more advanced study of the interplay between QCs and entanglement, see Ref.~\refcite{genio,streltsov}.

\subsection{Extension to  $2\otimes d$ systems}

Here we address the problem of measuring bipartite QCs for states of $2\otimes d$ systems, where subsystem $A$ is the qubit.  A generalization of geometric discord to catch bipartite QCs in such a case  has been derived (for finite $d$) in Ref.~\refcite{rau}.  Its expression is the very same as  Eq.~(\ref{eq1}). The Bloch-Fano form for the state is
 \begin{eqnarray}\label{ext}
  \rho &=& \frac 1{2d}\left(\mathbb{I}_{2d}+\sum_{i=1}^3 x_i\sigma_i \otimes \mathbb{I}_{d} +\sum_{j=1}^{d^2-1} y_j \mathbb{I}_{2}\otimes \tau_j+\sum_{i=1}^3\sum_{j=1}^{d^2-1} t_{ij} \sigma _i\otimes\tau_j\right),
 \end{eqnarray}
 where we can pick the generalized and normalized Gell-Mann matrices as  basis $\{\tau_j\}$ of the $d$-dimensional subsystem $B$. Obviously, $\{y_j\}$ is now a $d$-dimensional vector and $t$ is a $3\times d$ correlation matrix.
One can notice that the matrix $S=\frac{1}{2d}(X+T)$ has still  $3\times 3$ dimension, thus its characteristic equation remains a cubic and we can repeat all the previous steps to write closed expressions for $D_G$ and $Q$ formally equivalent to Eq.(\ref{dg},\ref{q}). The procedure can be extended to $d=\infty$ according to the prescription of Ref.~\refcite{rspibrido}.
\subsection{DQC1 calculations}
\begin{figure}[bt]
\begin{eqnarray*}
 \Qcircuit @C=1.4em @R=1.2em
{
\lstick{\frac 12(\mathbb{I}_2+\mu \sigma_3)} &  \gate{H} &  \ctrl{1}  &\meter \\
\lstick{\mathbb{I}_8/8} &\qw & \gate{U} &\qw\\
}
\end{eqnarray*}
\vspace*{8pt}
\caption{DQC1 model with a three-qubit state in a maximally mixed state and an attached ancilla of purity $\mu$. Measuring $\sigma_1, \sigma_2$ on the ancilla returns  the real and imaginary part of $\text{Tr}[U]$,}
\label{nm}
\end{figure}
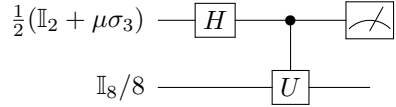

To conclude this section, we include a simple but meaningful case study to showcase a comparison of QCs measures.
The DQC1 algorithm, introduced in Ref.~\refcite{laf1}, is a non-universal quantum computing protocol estimating the trace of a $n$-qubit unitary matrix $U$. Ideally suited for an NMR setting, it provides an exponential speed up compared to the best known classical algorithm for such a specific task \cite{dattabarbieriaustinchaves,white,laf2,ginanuovo}. The surprising feature of the DQC1 is that this enhancement in the performance is obtained despite a negligible amount of entanglement created during the computation. In particular, an ancillary qubit (Alice's subsystem) is initially in a state with arbitrary polarization $\mu$, i.e., $\rho^{in}_A=\frac12(\mathbb{I}_2+\mu \sigma_3)$, while Bob amounts to an $n$-qubit maximally mixed state, i.e.,  $\rho_B^{in}=\frac1{2^n}\mathbb{I}_{2^n}$. A four-qubit  implementation (the ancilla $A$ {\it vs} $n=3$ qubits) has been recently investigated experimentally in Ref.~\refcite{laf2,ginanuovo}. Specifically, the designed unitary gate is $U=(a,a,b,1,a,b,1,1)$, with $a=-(e^{-i 3\pi/5})^4, b=(e^{-i 3\pi/5})^8$. An approximation of  the Jones polynomials has been realized by this setting \cite{jones}. The evaluation of $\text{Tr}[U]$ runs in the following way: given an initial uncorrelated state $\rho^{in}=\rho^{in}_A\otimes\rho^{in}_B$, referring to the scheme of Fig.~\ref{nm}, the protocol returns the final state
\begin{eqnarray}
\rho^{out}&=&(U_H\otimes\mathbb{I}_B)U\rho^{in}U^{\dagger}(U_H\otimes\mathbb{I}_B)^{\dagger}\nonumber\\
&=&\frac1{16}\left(
\begin{array}{c|c}
 \mathbb{I}_{8} & \mu U^{\dagger}  \\ \hline
 \mu U &   \mathbb{I}_8 \\
\end{array}
\right).
  \end{eqnarray}
where we denote by $U_H$ an Hadamard gate to be performed on Alice side. It is straightforward to see that the output state for Alice is
  \begin{eqnarray}
  \rho_A^{out}&=&\frac1{2}\left(
\begin{array}{cc}
 1 & \frac{\mu}{8} \text{Tr}[U^{\dagger}]  \\
 \frac{\mu}{8} \text{Tr}[U] &   1 \\
\end{array}
\right).
  \end{eqnarray}
Thus measurements of the ancilla polarization yield an estimation of the trace of the unitary matrix: $\langle\sigma_1\rangle_{\rho_A^{out}}=\text{Re}\left\{\text{Tr}[U]/8\right\}, \langle\sigma_2\rangle_{\rho_A^{out}}=\text{Im}\left\{\text{Tr}[U]/8\right\}$. \par
 Let us turn our attention to the correlations in the final state of the computation. The entanglement between the ancilla and the three-qubit register is manifestly negligible and does not scale with the number of qubits involved. On the same hand, we can easily study QCs across this bipartition. In particular, we compare the behaviour of geometric QCs measures, i.e., the geometric discord $D_G$ and the lower bound $Q$, with the entropic discord ${\cal D}$ defined in Refs.~\refcite{OZ,HV}.  The former measures are easily calculated from Eq.~(\ref{dg},\ref{q}), while for the latter one we retrieve the approximated expression for the output states of the DQC1 model calculated in Ref.~\refcite{altremisure}:
    \begin{figure}[bt]
\centerline{\psfig{file=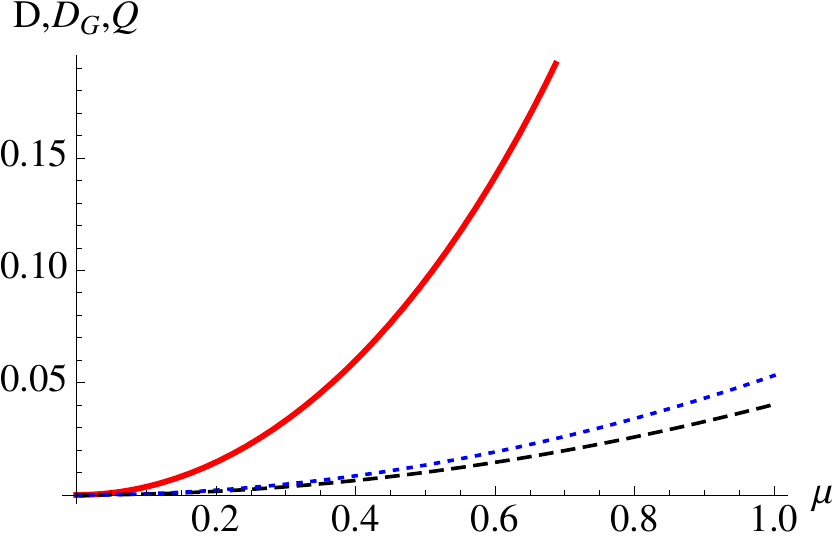,width=8cm}}
\vspace*{8pt}
\caption{Bipartite quantum correlations for the final state in the four-qubit DQC1 model discussed in the text, as measured by ${\cal D}$ (red continuous line), $D_G$ (blue dotted line) and $Q$ (black dashed line) as functions of the initial ancilla polarization $\mu$. All the plotted quantities are dimensionless.}
 \label{QvsD}
\end{figure}
 \begin{eqnarray}
 D_G&=&0.0531325 \mu^2\nonumber\\
 Q&=&0.0402856 \mu^2\nonumber\\
 {\cal D}&=&2-h_2\left(\frac{1-\mu}2\right)-\log_2[1+\sqrt{1-\mu^2}]-(1-\sqrt{1-\mu^2})\log_2[e],
 \end{eqnarray}
 where $h_2$ is the binary  Shannon entropy. Surprisingly, the geometric discord  of the DQC1 output state can be directly linked to the trace of the square of the unitary matrix (see Ref.~\refcite{ginanuovo} for details). In Fig.~\ref{QvsD} we study the behaviour of QCs measured by ${\cal D},D_G, Q$ by varying the purity of the ancilla in the initial state. As expected, the amount of bipartite QCs between the ancilla and the three-qubit register seems to be an efficient figure of merit for the efficiency of the protocol, as all the considered QCs measures are monotonically increasing with the initial purity of the ancilla.

 \section{Experimental implementation: a theoretical point of view}\label{three}

\subsection{Nuclear Magnetic Resonance (NMR) setting}
For this section, general references are Refs.~\refcite{brazirev,abra}. In NMR experiments, quantum states are realized by spinorial configurations of magnetic nuclei.  At room temperature, the $n$-qubit state of a NMR system is given by
\begin{eqnarray}
\rho=\frac{1}{2^n}\mathbb{I}_{2^n}+\epsilon\Delta\rho,
\end{eqnarray}
where $\epsilon=\hbar\omega_L/2^n K_BT\sim10^{-5}$ is the average thermal energy.  Every manipulation is implemented by varying the deviation matrix $\Delta\rho$, which carries the actual information content of the state. \par
A peculiarity of the NMR setting is that there is negligible entanglement in the produced states. In spite of that, several quantum computational tasks  have been satisfyingly studied and implemented by means of such a technique. It is supposed that QCs could be the key resources for the supraclassical performances in NMR environments. Indeed, the DQC1 model of computation we treated in the previous section was designed thinking about NMR implementation of quantum information processing \cite{laf1}. In any case, it appears a well suited ground for investigating QCs potentialities, and definitely this is not a coincidence. Measures of QCs such as geometric discord and $Q$ are the easiest to manage at theoretical level, built by considering the density matrix of the state in the Bloch-Fano picture.  This theoretical framework was just introduced in the Refs.~\refcite{bloch,fano} for efficiently describing the resonance of magnetic nuclei under the influence of an external magnetic field. Thus, what seems  at first sight a merely formal coincidence, underlines in fact a privileged interweaving between geometric quantification of QCs and NMR techniques. An overview of the recent studies of QCs in this setting can be found in Ref~\refcite{brazirev}.\par
In the NMR context, performing global and local spin measurements is the most convenient method for gaining information about a state. Tomography would definitely require the spin measurements necessary to retrieve all the $R_{ij}=\text{Tr}[(\sigma_i\otimes\tau_j)\rho]$ coefficients, and thus the state. On the other hand, evaluating  geometric discord $D_G$ and the lower bound $Q$, by definition, does not entail to know anything about Bob's subsystem, i.e., we can drop the $d^2-1$ measurements related to the Bloch vector $\vec{y}$, that is $y_j=\text{Tr}[(\mathbb{I}_2\otimes\tau_j)\rho]$. Therefore, denoting by $O_{i}$ the observables to be linked with the QCs measures, one can write
\begin{eqnarray}
\langle O_i^{\text{NMR}}\rangle&=&\text{Tr}[\sigma_{\nu}\otimes\tau_{\lambda}\rho], \nu=1,\ldots,3,; \lambda=0,\ldots, d^2-1\nonumber\\
D_G&=&f[\langle O_i^{\text{NMR}}\rangle]\nonumber\\
Q&=&\tilde{f}[\langle O_i^{\text{NMR}}\rangle].
\end{eqnarray}
    Thus, it is relatively easy to quantify QCs in NMR setting  \cite{laf2,ginanuovo,mazi,brazi2}.  An experimental trick for further simplifying the considered procedure allows to restate global spin measurements as local ones on Alice only, by just running a global operation on the state before the measurements \cite{brazirev}. For each particular measurement, a related global operation is selected. Specifically,
   \begin{eqnarray}
&\xi_{\nu}=&U_{\text{CNOT}}R_{A}(\phi, n)\otimes R_{B}(\phi, n)\rho (U_{\text{CNOT}}R_{A}(\phi, n)\otimes R_{B}(\phi, n))^{\dagger}\nonumber\\
   &\text{Tr}&[(\sigma_{\nu}\otimes\sigma_{\lambda})\rho]=\text{Tr}[(\sigma_{\nu}\otimes\mathbb{I}_d)\xi_{\nu}],
   \end{eqnarray}
 where we have applied local rotations $R$ by an angle $\phi$ about some direction $n$, both dependent on the specific $\sigma_{\nu}$ to be evaluated, and subsequently a CNOT gate with Alice being the control qubit.
   One could maintain that when $d>2$ the global spin measurements, i.e., estimations of the expectation values of $\sigma_{\nu}\otimes\tau_{\lambda}$ as introduced in Eq.~(\ref{ext}), seem extremely intricate, and the realization of the global rotation might be beyond the current technological possibilities. In such a case, at least for the paradigmatic instance in which Bob is a $n$-qubit subsystem ($d$ is even), we can pick, as basis $\{\tau_j\}$ for the  $d$-dimensional subsystem,  the tensor products of Pauli matrices
  \begin{eqnarray}\label{pauli}
  \{\tau_\lambda\}=\{\mathbb{I}_{d}, \sigma_1\otimes\mathbb{I}_{d-2},\sigma_2\otimes\mathbb{I}_{d-2},\sigma_3\otimes\mathbb{I}_{d-2}, \mathbb{I}_2\otimes\sigma_1\otimes\mathbb{I}_{d-4},\ldots,\sigma_3\otimes\ldots\otimes\sigma_3\}
  \end{eqnarray}
  reducing the  detection of $t_{ij}$ to local spin measurements on single qubits only.
In summary, for the NMR set up, the QCs quantification, by both geometric discord and $Q$, demands $3d^2$ measurements, against the $4 d^2-1$ required by full state reconstruction. Indeed, we are exempt from making local spin measurements on Bob's side.

\subsection{Quantum Optics}
As mentioned in the introduction, the estimation of functionals of density matrix elements  $\rho_{ij}$  has been vastly investigated  by quantum optical setting. Some devices for the evaluation of meaningful quantities, for example the purity of  the state, have been built having as toolbox just the very basic principles of quantum computation. For a broad perspective on theoretical and experimental  features the reader should refer to Refs.~ \refcite{paisa12}--\refcite{horo0}.\par
It has been proven that any function of the density matrix entries can be expressed in terms of observables represented by hermitian unitary operators, and one can develop the experimental architecture to perform effective estimations in real world. It is important to stress that limits set by the actual technology could prevent from implementing what has been successfully designed. So, from the very beginning, we look for a QCs measure \emph{really} accessible to experimentalists. \par
Let us consider the specific case study of the implementation of $Q$ for a two-qubit state. The task, as clearly expressed in Eq.~(\ref{q}), is to recast the quantities $\text{Tr}[S]$ and $\text{Tr}[S^2]$ in terms of observables.\par
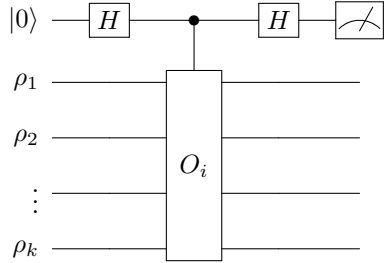
\begin{figure}[bt]
\begin{eqnarray*}
 \Qcircuit @C=1.4em @R=1.2em
{
\lstick{\ket{0}} &  \gate{H} &  \ctrl{1} &  \gate{H} &\meter \\
\lstick{\rho_1} &\qw & \multigate{3}{O_i} & \qw &\qw\\
\lstick{\rho_2} &\qw & \ghost{O_i} & \qw &  \qw\\
\lstick{\vdots} &\qw & \ghost{O_i} & \qw & \qw \\
\lstick{\rho_k} &\qw & \ghost{O_i} & \qw &  \qw
}
\end{eqnarray*}
  \caption{Circuit estimating $\text{Tr}[O_i \rho_1\otimes \rho_2\otimes\ldots\otimes\rho_k]= v$.  A Hadamard gate $H$ is applied to the ancillary qubit, followed by a controlled-$O_i$ gate acting on the overlap of states and then another Hadamard one. Then, a measurement in the computational basis returns the visibility $v$, i.e., the expectation value of the operator $O_i^{\text{OPT}}$.}
    \label{magic}
\end{figure}

On can see that
\begin{equation}
\text{Tr}[S]=\frac14(\text{Tr}[X]+\text{Tr}[T]),\text{Tr}[S^2]=\frac 1{16}(\text{Tr}[X^2]+\text{Tr}[T^2]+2 \text{Tr}[XT])\,.\end{equation}
 After some algebra, one obtains
 \begin{eqnarray}\label{nqubit}
\text{Tr} [X]&=& 2 \text{Tr}[\rho_A^2]-1 \nonumber\\
\text{Tr} [T]&=&  4(\text{Tr}[ \rho^2]-\text{Tr}[ \rho_A^2]/2-\text{Tr}[ \rho_B^2]/2)+1\nonumber\\
\text{Tr}[X^2]&=& (2\text{Tr}[\rho_A^2]-1)^2\nonumber\\
\text{Tr}[XT]&=&-1+4 \text{Tr}[\rho^2](-1+\text{Tr}[\rho_A^2])+4 \text{Tr}[\rho_A^2]-4\text{Tr}[\rho_A^2]^2+2\text{Tr}[\rho_B^2]\nonumber\\
&+&8 \text{Tr}[\rho (\rho_A\otimes \mathbb{I}_{2}) \rho(\rho_A\otimes \mathbb{I}_2)]-8 \text{Tr}[\rho (\rho_A^2\otimes \rho_B)] \nonumber\\
\text{Tr}[T^2]&=& -32(\text{Tr}[\varsigma^4]+ \text{Tr}[\varsigma^3])+3( \text{Tr}[T]^2/2 -\text{Tr}[T]-1/2 ),
\end{eqnarray}
where $\varsigma =\rho - (\rho_A\otimes\mathbb{I}_2)/2- (\mathbb{I}_2\otimes\rho_B)/2$. Consequently, we can write $Q$ in terms of traces of multicopies of the global and marginal density matrices and their overlaps. In particular:
\begin{eqnarray}\label{s}
\text{Tr}[S]&=&\text{Tr}[\rho^2]-\text{Tr}[\rho_B^2]/2\nonumber\\
\text{Tr}[S^2]&=&\frac14(-2-8\text{Tr}[\rho^4]+8\text{Tr}[\rho^3]+ 6\text{Tr}[\rho^2]^2\nonumber\\
&-&2 \text{Tr}[\rho^2](5+\text{Tr}[\rho_B^2])-2 \text{Tr}[\rho_A^2]^2+10\text{Tr}[\rho_A^2]\nonumber\\
&-&\text{Tr}[\rho_B^2]^2+12\text{Tr}[\rho_B^2]-6\text{Tr}[\rho_A^2]\text{Tr}[\rho_B^2]\nonumber\\
&+&4\text{Tr}[\rho (\mathbb{I}_2\otimes\rho_B)\rho(\mathbb{I}_2\otimes\rho_B)]-24\text{Tr}[\rho (\rho_A\otimes\rho_B)]\nonumber\\
&+&8\text{Tr}[\rho (\rho_A\otimes\mathbb{I}_2)\rho(\rho_A\otimes\mathbb{I}_2)]+8\text{Tr}[\rho^2(\rho_A\otimes\rho_B)]) .
\end{eqnarray}
Therefore, $Q$ can be recast as a functional of polynomials (of up to the fourth order) of the density matrix elements,  specifically traces of matrix powers and overlaps. We identify nine independent terms in the expressions of Eq.{~(\ref{s})}. Inspired by the historical lesson of nontomographic entanglement detection \cite{mint1,mintexp}, we associate to them the expectation values of the operators $\{O_i^{\text{OPT}}\}_{i=1}^9$. Explicitly, the $O_i^{OPT}$ are swap/shift operators $V^k$ acting on $k$ ($k \le 4$)  copies of the global and/or marginal density matrices and related overlaps.  For a density matrix $\rho$  it holds that $\text{Tr}[\rho^k]=\text{Tr}[V^k \rho^{\otimes k}]$ \cite{paisa12,paz,brun,ekert,winter,horo0}, where $V^k$ is the shift operator, $V^k|\psi_1\psi_2\ldots\psi_k\rangle=|\psi_k\psi_1\ldots\psi_{k-1}\rangle$. Also, for two unknown states $\rho_1,\rho_2$, it has been proven that $ \text{Tr}[V^2 \rho_1\otimes \rho_2]=\text{Tr}[\rho_1\rho_2]$ \cite{ekert,filip1,filip2}.  More generally, we have $\text{Tr}[\rho_1\rho_2\ldots\rho_k]=\text{Tr}[V^k \rho_1\otimes \rho_2\otimes\ldots\otimes\rho_k]$. We briefly present a proof of the last statement, see also Ref.~\refcite{linden} for a more elegant treatment.
One can see that $\text{Tr}[\rho_1\rho_2\ldots\rho_k]=\text{Tr}[V^k \rho_1\otimes \rho_2\otimes\ldots\otimes\rho_k]$ by expanding the left-hand term in the equation as follows:
\begin{eqnarray}
\text{Tr}[\rho_1\rho_2\ldots\rho_k]&=&\sum_{i_1j_1\ldots i_kj_k}\rho^{i_1}_{1\  j_1}\rho^{i_2}_{2\ j_2}\ldots \rho^{i_{k-1}}_{k-1\ j_{k-1}}\rho^{i_k}_{k\ j_k} \delta_{i_2}^{j_1}\delta_{ i_3}^{j_2}\ldots\delta_{i_k}^{j_{k-1}}\delta^{j_k}_{ i_1}\nonumber\\
&=&\sum_{i_1\ldots i_k}\rho^{i_1}_{1\ i_2}\rho^{i_2}_{2\ i_3}\ldots \rho^{i_{k-1}}_{k\ i_k}\rho^{i_k}_{k\ i_1}.
\end{eqnarray}
Denoting by $\{|i\rangle\}$ a Hilbert space basis, $\rho_{1\ldots k}=\otimes_i \rho_i$,  and building the shift operator as chain of swaps $V^2=\sum_{i_A i_B j_A j_B}|i_A j_B\rangle\langle j_A i_B|$, one has \begin{eqnarray}\label{generalisedeckert}
V^k \rho_1\otimes \rho_2\otimes\ldots\otimes\rho_k&=&\sum_{i_1j_1\ldots i_kj_k} \rho^{i_1\ldots i_k}_{1\ldots k  j_1\ldots j_k}|i_1j_k\ldots i_{k-1}\rangle\langle j_1 i_{k}\ldots j_{k-1}|\ldots\nonumber\\
\ldots&&|i_1\ldots i_{k-1}j_k\rangle\langle j_1\ldots j_{k-1} i_{k} |i_1\ldots i_{k-1} i_k\rangle\langle j_1\ldots j_{k-1} j_k|,\nonumber\\
\
\end{eqnarray}
 while the right-hand  term of the initial relation takes the form
\begin{eqnarray}
\text{Tr}\left[V^k \rho_1\otimes \rho_2\otimes\ldots\otimes\rho_k\right]&=&\sum_{i_1\ldots i_k}\rho^{i_1\ldots i_k}_{1\ldots k\ i_2\ldots i_{1}}\delta^{i_1}_{i_k}\nonumber\\
&=& \sum_{i_1\ldots i_k}\rho^{i_1}_{1\ i_2}\rho^{i_2}_{2\ i_3}\ldots \rho^{i_{k-1}}_{k\ i_k}\rho^{i_k}_{k\ i_1},
\end{eqnarray}
thus the assertion is proven. For example, $\langle O_1^{\text{OPT}}\rangle=\text{Tr}[\rho^4]=\text{Tr}[V^4\rho^{\otimes 4}]$, and so forth for the other terms.  All the quantum circuits to be implemented for estimating $\langle O_i^{\text{OPT}}\rangle$ (by inserting  an ancillary qubit) have the same architecture, which is depicted in  Fig.\ref{magic}.
A Mach-Zender  interferometer is modified by a controlled-$O_i$ gate. At the end of the routine, the visibility $v$ related to the interference fringes yields the expectation value of the operator $O_i^{\text{OPT}}$ on the general dummy overlap $\rho_1\otimes \rho_2\otimes\ldots\otimes\rho_k$, as one can write $\text{Tr}[O_i^{\text{OPT}} \rho_1\otimes \rho_2\otimes\ldots\otimes\rho_k]= v$.
Hence, to quantify the degree of QCs for an arbitrary two-qubit  state $\rho$, this method requires $9$ independent measurements instead of the $15$ necessary for tomography \cite{prl}. \par
Anyway we can indeed do better than this. In particular, also looking at Ref.~\refcite{cina}, we may appreciably  reduce the number of measurements and  the complexity of the setting. We observe that $V^2=V_{ij}=\sum_{ij}|ij\rangle \langle ji|=\frac{1}{2}(\mathbb{I}_{4}+\sum_k\sigma_k\otimes\sigma_k)$  (note that the $V$ defined in Ref.~\refcite{cina} is twice ours). Then, let us introduce the following quantities
\begin{eqnarray}
c_1&=&\text{Tr}[(P_{A_1A_2}^-\otimes P_{B_1B_2}^-)(\rho^{\otimes 2})]\nonumber\\
c_2&=&\text{Tr}[(P_{A_1A_2}^-\otimes \mathbb{I}_{B_1B_2})(\rho^{\otimes 2})]\nonumber\\
c_3&=&\text{Tr}[(\mathbb{I}_{A_1A_2}\otimes P_{B_1B_2}^-)(\rho^{\otimes 2})]\nonumber\\
c_4&=&\text{Tr}[(P_{A_1A_4}^-\otimes P_{A_2A_3}^-\otimes P_{B_1B_2}^-\otimes P_{B_3B_4}^-)(\rho^{\otimes 4})]\nonumber\\
c_5&=&\text{Tr}[(P_{A_1A_4}^-\otimes \mathbb{I}_{A_2A_3}\otimes P_{B_1B_2}^-\otimes P_{B_3B_4}^-)(\rho^{\otimes 4})]\nonumber\\
c_6&=&\text{Tr}[(P_{A_1A_4}^-\otimes P_{A_2A_3}^-\otimes P_{B_1B_2}^-\otimes \mathbb{I}_{B_3B_4})(\rho^{\otimes 4})]\nonumber\\
c_7&=&\text{Tr}[(\mathbb{I}_{A_1A_4}\otimes P_{A_2A_3}^-\otimes P_{B_1B_2}^-\otimes \mathbb{I}_{B_3B_4}^-)(\rho^{\otimes 4})],
\end{eqnarray}
 where $P_{ij}^-=\frac{1}{2}(1-V_{ij})$ is the projector on the antisymmetric subspace for a two-qubit state. Evaluating $c_i$ is equivalent to measuring the quantities
\begin{eqnarray}
d_1&=&\text{Tr}[(V_{A_1A_2}\otimes V_{B_1B_2})(\rho^{\otimes 2})]\nonumber\\
d_2&=&\text{Tr}[(\mathbb{I}_{A_1A_2}\otimes V_{B_1B_2})(\rho^{\otimes 2})]\nonumber\\
d_3&=&\text{Tr}[(\mathbb{I}_{A_1A_4}\otimes V_{A_2A_3}\otimes V_{B_1B_2}\otimes V_{B_3B_4})(\rho^{\otimes 4})]\nonumber\\
d_4&=&\text{Tr}[(V_{A_1A_4}\otimes V_{A_2A_3}\otimes V_{B_1B_2}\otimes V_{B_3B_4})(\rho^{\otimes 4})].
\end{eqnarray}
More important, one can see that
\begin{eqnarray}
\text{Tr}[S]&=&4 c_1-2 c_2-c_3+\frac12= d_1 - \frac12 d_2 \nonumber\\
\text{Tr}[S^2]&=& 16 c_4 + 8(c_7 - c_5-2 c_6)+c_3^2+4 c_2^2-c_3-2 c_2+\frac12\nonumber\\
&=&d_4 - d_3 + \frac1{4} d_2^2.
\end{eqnarray}
Therefore, $Q$ could be detected by evaluating seven projective or even just four swap measurements:
\begin{eqnarray}
Q&=&g[\langle O_i^{\text{OPT}}\rangle]\nonumber\\
\{O_i^{\text{OPT}}\}&=&\{c_i\}\ \text{or}\ \{d_i\}.
\end{eqnarray}
 We remark that  geometric discord  would require a rather more complex expression in terms of overlaps or alternatively measurements over six copies of the state \cite{cina}, entailing by far a harder implementation.\par
 Now, let us have a look at the extension to the $2\otimes d$ case. We can arguably say that the very same expressions hold, at least at formal level. Clearly, we have to generalize the swap and the projectors to arbitrary finite dimension.  A state of a $d$-dimensional system reads  $\rho=\frac1d(\mathbb{I}_d+\sum_i x_i\tau_i)$, implying  $\text{Tr}[\rho^2]=\frac1d(1+|\vec{x}|^2)$. Thus, in the most general fashion one obtains
\begin{eqnarray}
V=\frac1d(\mathbb{I}_{d^2}+\sum_i \tau_i\otimes\tau_i),
\end{eqnarray}
and consequently
\begin{eqnarray}
P^-=\frac1{2d}((d-1)\mathbb{I}_{d^2}-\sum_i \tau_i\otimes\tau_i),
\end{eqnarray}
where the $\tau_i$s reduce to Pauli matrices in $d=2$.  The optical implementation of projectors on $P_{B_iB_j}$, i.e. multiqubit projectors, is more complicated than the two-qubit case, see e.g.~Ref.~\refcite{ent}.  However, the method presented in Ref.~\refcite{cinaexp} can be extended to arbitrary dimensions without terribly increasing the complexity of the experimental setting. More precisely, the necessary number of optical devices should increase polynomially with $d$, not exponentially. See also Ref.~\refcite{2cinaexp} for more a detailed analysis. One can appreciate that the number of measurements is independent of Bob's dimension.
Anyway, it could be pointed out that swaps and projectors on large dimensional systems seem of hard implementation. However, such a question can be easily overcome in two ways. First, as already done for the NMR setting, we can restrict to the case of even $d$ and pick for Bob's subsystem the basis considered in Eq.~(\ref{pauli}),  such that  $V,P^-$ can be easily rewritten in terms of two-qubit operators. In such a case, the number of measurements would increase linearly with $d$ not compromising the scalability of the protocol.  Alternatively, we can directly rewrite $d$-dimensional swaps/projectors as matrix products of their two-qubit versions. We leave for future investigations an extensive treatment of this issue.

\section{Quantum correlation dynamics in open systems}\label{four}

\subsection{Overview}

In this section we will have a look at the properties of QCs, and in particular of the quantifier $Q$, in the context of dynamical open quantum systems. The general aim of the study of open quantum systems is to provide tools to characterize the dynamics of the state $\rho_S(t)$ of a system of interest $S$ when it interacts with an environment $E$ \cite{BrePet,Weiss}. One of the possible ways to solve this problem is to derive an equation of motion for $\rho_S(t)$, which is usually obtained starting from the Von Neumann equation for the global state of system and the environment, then tracing out the degrees of freedom of the environment. The result is an operator equation for the state of the system only, whose structure depends however on the system-environment interaction and on all the physical parameters involved, e.g. frequencies of the environmental degrees of freedom or coupling strengths.

The dynamics of correlations in open system contexts, e.g. entanglement or more general QCs, has been analyzed deeply and broadly in the last years \cite{YuEb,Prau,ManParOli,Paz,mazzola3,Vasile1,DiscQubit,Maziero,Vasile2,fanchini,PazDisc}. A key effect of the interaction with an environment is, for instance, the loss of bipartite correlations for long times, with the consequence of a final equilibrium state not presenting
typical signatures of quantumness. While quantum features are lost during the whole evolution, it is not uncommon to observe a partial restoration of such properties for certain time intervals of the dynamics. These \emph{non-Markovian} effects are interpreted as a consequence of memory effects of the environment, capable of restoring quantum coherence or correlations for short periods of time. Such a behavior usually is manifested in the very early stages of the dynamics (non-Markovian time scales) at least for weakly coupled systems, e.g. optical systems, and therefore it is experimentally challenging to observe. A recent successful attempt is reported in Ref.~\refcite{breuerexp} where the authors make use of the measure introduced in Ref.~\refcite{breuerthe}.
Another interesting result appears in Ref.~\refcite{LauraPRL}, where the authors show how QCs measured by quantum discord are frozen in the early stages of the dynamics. Such a transition from a classical to a quantum decoherence appears qualitatively different if other measures of quantum correlations are employed, e.g. geometric discord \cite{dakic}. As we will review in Section \ref{four1}, geometric discord does not remain constant at any time, but a change of dynamical behavior is anyway observed at the transition time evaluated in Ref.~\refcite{LauraPRL}.

The main point of interest of this section is to establish whether the approximate value of geometric discord captured by the quantifier $Q$ is faithful during the open system dynamics of the system of interest. In the two examples we consider in the next section we will see that, since the state changes in time according to the open system dynamics, the difference between $D_G$ and $Q$ will be in general function of time. Differences in the dynamics of $D_G$ and $Q$ are clearly due to the peculiar evolution of the parameter $\theta$ defined in Eq. \eqref{10}.

\subsection{Independent environments: non-Markovian case}

The first open system model we introduce is a system made of two identical non-interacting qubits, each coupled with its own bosonic reservoir at zero temperature. The total Hamiltonian $H$ is then written as a sum of two terms, $H_1$ and $H_2$ having the form
\begin{equation}
H_i=\omega_0\sigma^{(i)}_+\sigma^{(i)}_- +\sum_k\omega_k b^{(i)\dag}_k b^{(i)}_k +\sum_k g_k (\sigma^{(i)}_+ b^{(i)}_k + \sigma^{(i)}_- b^{(i)\dag}_k)
\end{equation}
where $\omega_0$ is the frequency of the qubits, the $\omega_k$ are the frequencies of the environmental oscillators, the $g_k$ quantify
the coupling of the oscillators with the $k$-th mode of the respective environment, $\sigma^{(i)}_+$ and $\sigma^{(i)}_-$ are the Pauli raising and lowering operators for the $i$-th oscillator, and, finally, the $b^{(i)}_k$ and $b^{(i)\dag}_k$ are the annihilation and creation operators for the mode $k$ of the $i$-th reservoir. From the previous definition it should be clear we are under the assumption that the environments have the same properties and are coupled with the same strength to the respective qubits. The properties of the reservoirs at zero temperature can be all condensed in the knowledge of their spectral function $J(\omega)$. Here we choose a Lorentian shaped spectral distribution of the form
\begin{equation}
J(\omega)=\frac{1}{2\pi}\frac{\gamma_0\lambda^2}{(\omega_0-\omega)^2+\lambda^2}
\end{equation}
where $\gamma_0$ is an overall coupling constant, $\lambda$ is the width of the spectral function and $\omega_0$ is the center of the spectrum, which is resonant to the free qubit frequencies.

\begin{figure}[bt]
\centering
\subfigure[]
{\includegraphics[width=6cm]{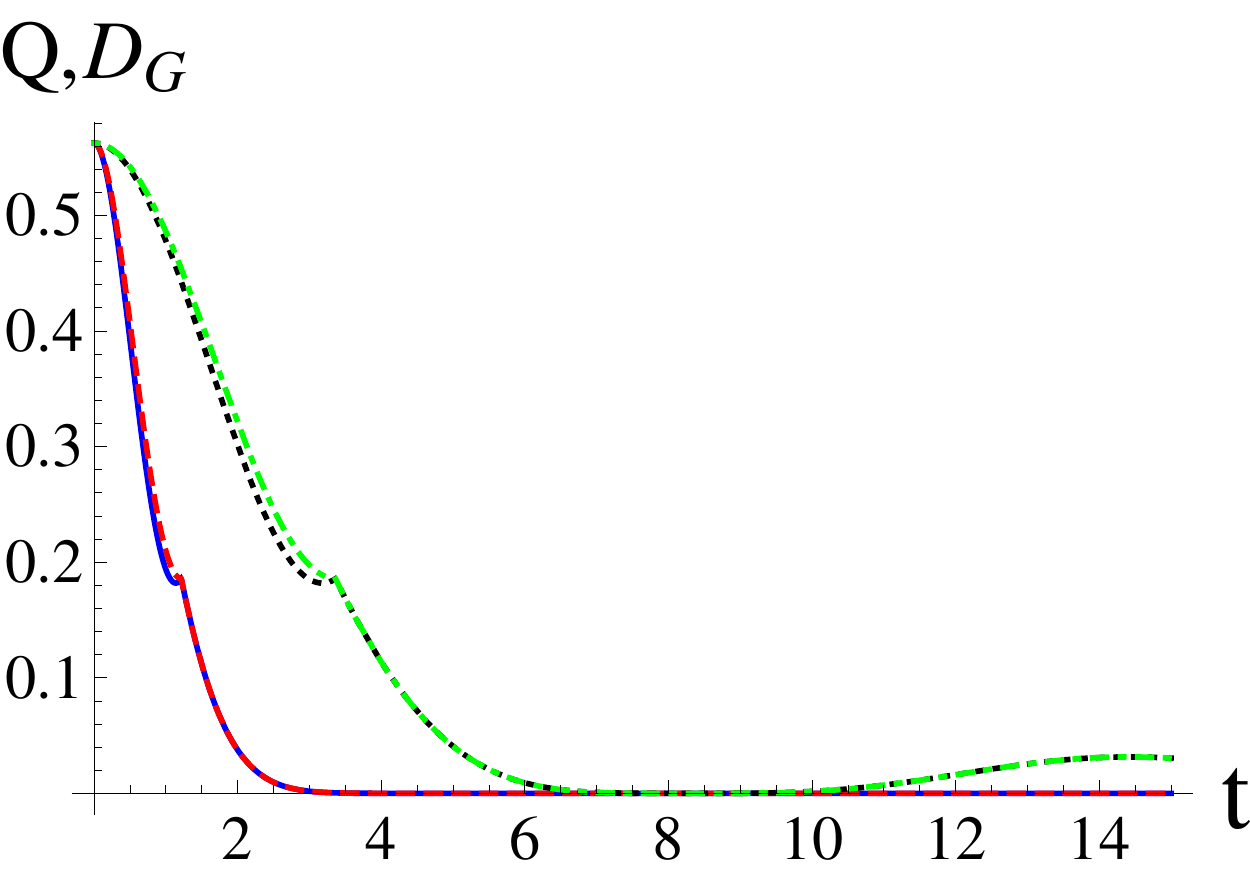}\label{Fig1A}}
\hspace*{0.2cm} \subfigure[]
{\includegraphics[width=6cm]{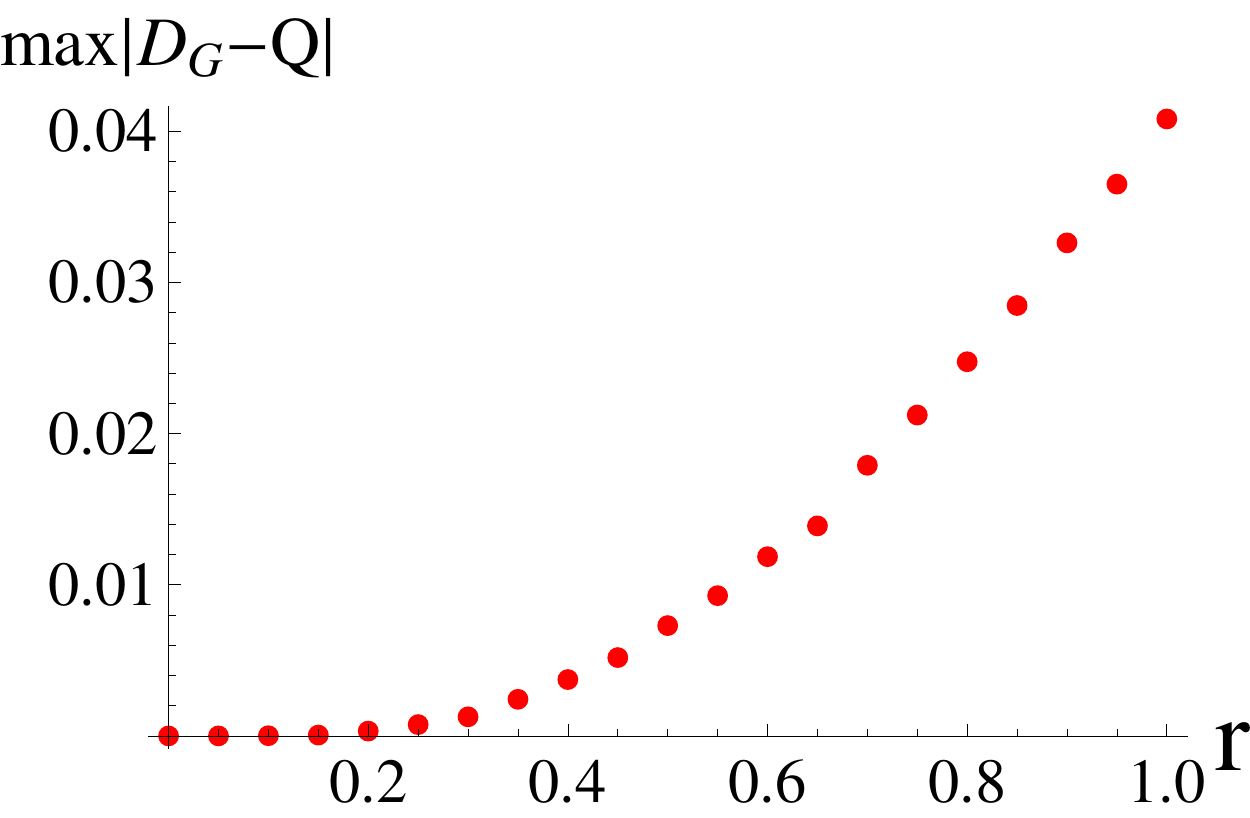}\label{Fig1B}}
\caption{(a) Correlation quantifier $Q$ and geometric discord $D_G$ for an initial two qubit Werner state with $r=3/4$ as a function of time $t$. Environment parameters $\gamma_0=1$, $\lambda=1$ (blue solid, $Q$ and red dashed $D_G$ lines) and $\lambda=0.1$ (black dotted, $Q$, and green dotted-dashed, $D_G$, lines). (b) Maximum difference between geometric discord $D_G(t)$ and quantifier $Q(t)$ for different initial Werner states (parameter $r$ is varied).}
\label{Fig1B}
\end{figure}

Since the total Hamiltonian is given by the sum of two commuting terms, for any initial state the dynamics of the system can be solved directly from the knowledge of the solution for a single qubit in its environment as proven in Ref.~\refcite{BelLoCom}. Given the density operator $\rho(0)$ at the initial time $t_0=0$, its density matrix elements in the canonical basis $\{\ket{1}=\ket{11},\ket{2}=\ket{10},\ket{3}=\ket{01},\ket{4}=\ket{00}\}$ at time $t$ are given by
\begin{equation}\label{SolInd}
\begin{array}{lll}
\rho_{11}(t)=\rho_{11}(0)P_t^2, &\qquad & \rho_{22}(t)=\rho_{22}(0)P_t+\rho_{11}(0)P_t(1-P_t),\\
\rho_{33}(t)=\rho_{33}(0)P_t+\rho_{11}(0)P_t(1-P_t),&\qquad&\rho_{44}(t)=1-\rho_{11}(t)-\rho_{22}(t)-\rho_{33}(t),\\
\rho_{12}(t)=\rho_{12}(0)P_t^{3/2},&\qquad& \rho_{13}(t)=\rho_{13}(0)P_t^{3/2},\\
\rho_{14}(t)=\rho_{14}(0)P_t,&\qquad& \rho_{23}(t)=\rho_{23}(0)P_t,\\
\rho_{24}(t)=P_t^{1/2}[\rho_{24}(0)+\rho_{13}(0)(1-P_t)]&\qquad& \rho_{34}(t)=P_t^{1/2}[\rho_{34}(0)+\rho_{12}(0)(1-P_t)].
\end{array}
\end{equation}
where we defined
\begin{equation}\label{EvoInd}
P_t=e^{-\lambda t}\biggl[\cos\Delta t+\frac{\lambda}{2\Delta}\sin\Delta t\biggl]
\end{equation}
with $\Delta=\sqrt{2\gamma_0\lambda-\lambda^2}/2$. Equation \eqref{EvoInd} contains all the information on the dynamics.
Since we have the expression for the state we can now apply the results of the previous sections and study the dynamics of the geometric discord $D_G$ \cite{dakic} and quantifier $Q$ \cite{prl} for any initial state. Here we consider the case of initial Werner states, having the form
\begin{equation}
\rho_{W}(r)=r\ket{+}\bra{+}+\frac{1-r}{4}\mathbb{I}_4=\frac{1}{4}\left(
       \begin{array}{cccc}
         1-r & 0 & 0 & 0 \\
         0 & 1+r & 2r & 0 \\
         0 & 2r & 1+r & 0 \\
         0 & 0 & 0 & 1-r \\
       \end{array}
     \right)
\end{equation}
where $\ket{+}=(\ket{2}+\ket{3})/\sqrt{2}$ is a maximally entangled Bell state. During the dynamics described by Eqs. \eqref{SolInd} and \eqref{EvoInd} the evolution brings the system through states for which the geometric discord $D_G$ and the quantifier $Q$ assume different values. Just as an example we show in Fig.~\ref{Fig1A}  the dynamics for an initial Werner state with $r=1/2$ as a function of $t$. The environment is characterized by $\gamma_0=1$, and two different values for the width, $\lambda=1$ (Markovian case) and $\lambda=0.1$ (non-Markovian case). In both situations, $D_G$ and $Q$ show the same qualitative dynamics (decay and oscillations). A small difference in the actual amount of correlations quantified appears in the first stage of the evolution. For completeness we show in Fig.~\ref{Fig1B} the maximum value of this difference, $\max_{t>0}[D_G(t,r)-Q(t,r)]$ as a function of the Werner parameter $0\leq r\leq 1$. The result clearly states that as the parameter $r$ grows, the quantifier $Q$ is a less tight bound.

\subsection{Depolarizing Markovian channel}\label{four1}

The second example we consider here is the case of two qubits evolving accordingly to two identical independent depolarizing channels \cite{LauraPRL}. The equation of motion for the single qubit channel has the form
\begin{equation}
\frac{d}{dt}\rho_{i}=\frac{\gamma}{2}(\sigma^{(j)}_i\rho_i\sigma^{(j)}_i-\rho_i)
\end{equation}
where $i=A,B$ is the qubit index, while $\sigma^{(1)}_i$, $\sigma^{(2)}_i$ and $\sigma^{(3)}_i$ are three Pauli matrices for the $i$-th
qubit. For more details the reader can refer to \refcite{LauraPRL} and references therein. We consider here only the phase flip channel case, i.e. $j=3$, and initial maximally mixed marginals states of the form
\begin{equation}\label{StateForm}
\rho_{AB}=\frac{1}{4}\biggl(\mathbb{I}_{AB}+\sum_{i=1}^3c_i(0)\sigma_i^A\sigma_i^B\biggl)
\end{equation}
with $0\leq |c_i|\leq 1$. The time evolution of the state is given by
\begin{equation}
\rho_{AB}(t)=\lambda^+_{\Psi}\ket{\Psi^+}\bra{\Psi^+}+
\lambda^+_{\Phi}\ket{\Phi^+}\bra{\Phi^+}+\lambda^-_{\Phi}\ket{\Phi^-}\bra{\Phi^-}
+\lambda^-_{\Psi}\ket{\Psi^-}\bra{\Psi^-}
\end{equation}
where $\ket{\Psi^+}=(\ket{00}\pm\ket{11})/\sqrt{2}$ and
$\ket{\Phi^+}=(\ket{01}\pm\ket{10})/\sqrt{2}$ are Bell states and
\begin{equation}\begin{split}
&\lambda^{\pm}_{\Psi}=[1\pm c_1(t)\mp c_2(t)+c_3(t)]/4\\
&\lambda^{\pm}_{\Phi}=[1\pm c_1(t)\pm c_2(t)-c_3(t)]/4
\end{split}\end{equation}
Finally we have $c_1(t)=c_1(0)\exp(-2\gamma t)$, $c_2(t)=c_2(0)\exp(-2\gamma t)$
and $c_3(t)=c_3(0)$.

\begin{figure}[bt]
\centering
\subfigure[]
{\includegraphics[width=6cm]{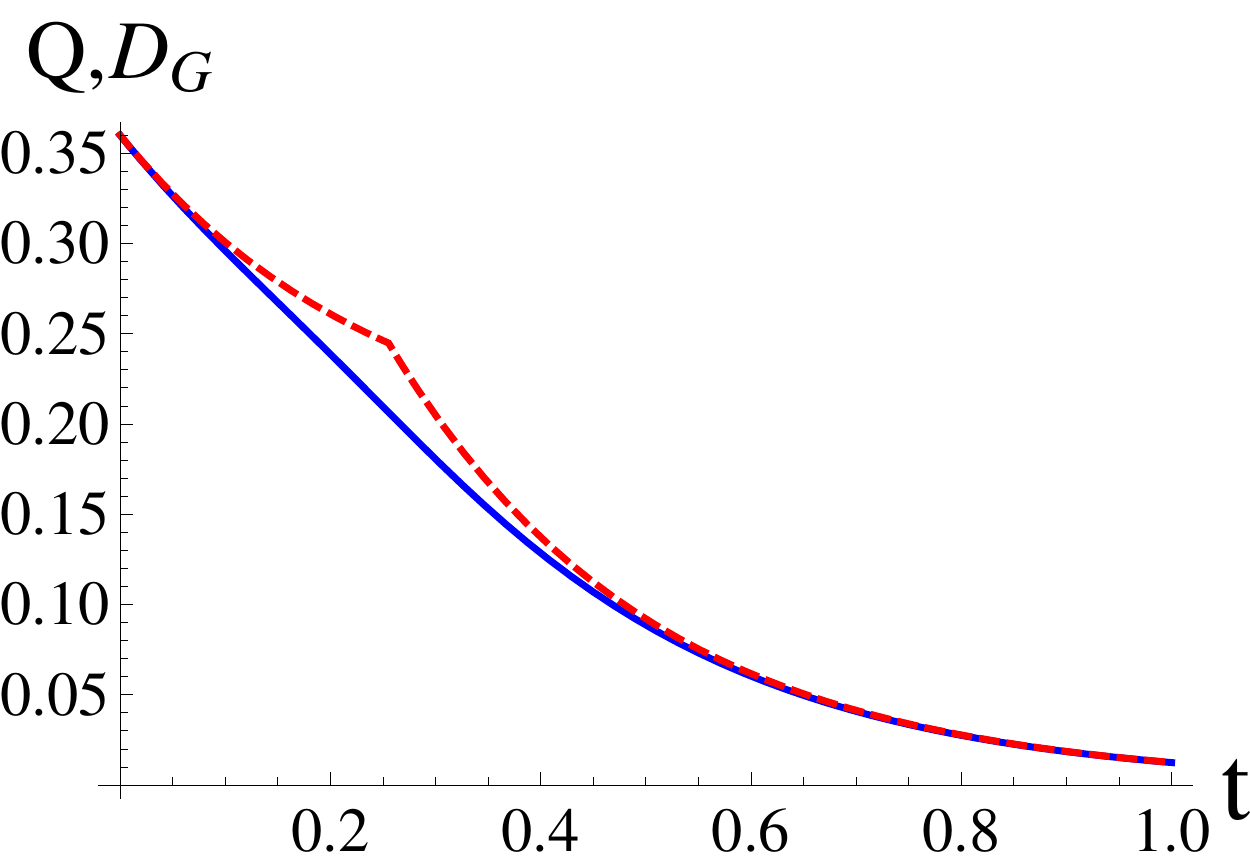}\label{Fig2A}}
\hspace*{0.2cm} \subfigure[]
{\includegraphics[width=6cm]{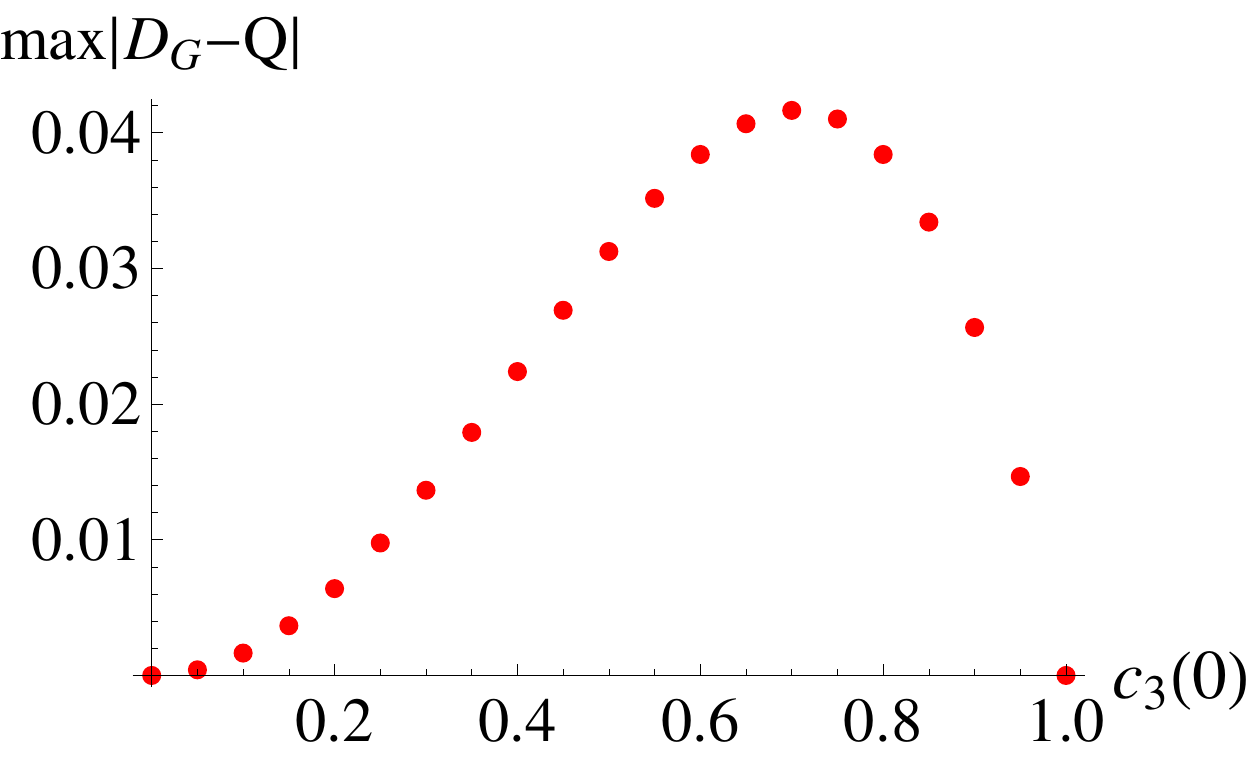}\label{Fig2B}}
\caption{(a) Time evolution of the correlation quantifier $Q$ (blue solid line) and geometric discord $D_G$ (red dashed line) for an initial two qubit state of the form \eqref{StateForm} with $c_1(0)=1$, $c_3(0)=-c_2(0)=0.6$ and $\gamma_0=1$. (b) Maximum difference between geometric discord $D_G(t)$ and quantifier $Q(t)$ for different initial states (parameters $c_3(0)=-c_2(0)$ are varied).}
\label{Fig2B}
\end{figure}

In Fig.~\ref{Fig2A} we plot the dynamics of geometric discord $D_G$ and the quantifier $Q$ for $\gamma=1$, $c_1(0)=1$ and $c_3(0)=-c_2(0)=0.6$. As one could expect, the sudden transition is witnessed only by the geometric discord, while the evolution of the quantifier does capture this dynamical feature. From Fig.~\ref{Fig2B} we show the maximum difference between geometric discord and $Q$. This maximum happens to be in correspondence with the transition point.

The results of this section show that, essentially, also in dynamical contexts (at least in these two examples shown) the correlation quantifier $Q$ behaves as a good tight lower bound to the actual geometric discord and therefore its use is encouraged also in the case of open system dynamics thanks to its improved accessibility. Some distinctive dynamical features are however not captured, and the origin of this has to be linked to the fact that, unlike $D_G$, $Q$ is defined without a minimization procedure over projective measurements.

\section{Conclusions}\label{con}
We discussed a proposal for quantifying  bipartite QCs in states of $2\otimes d$ dimensional composite systems at both theoretical and experimental level, building upon the results of Ref.~\refcite{prl}.  An analytical study of geometric discord led to define a tight meaningful lower bound $Q$ for two-qubit states, successively extended to $2\otimes d$ dimensions. From a theoretical point of view, $Q$ is faithful and easy to calculate, being characterized by a state-independent expression. On the experimental side, $Q$ is the friendliest measure we have so far, as its evaluation in laboratory appears to be in the reach of current technological levels and flexible enough to allow direct implementations in both NMR and optical settings. Indeed, $Q$  is just a function of polynomials of the density matrix entries, thus we can set a number of operators $\{O_i\}$ whose expectation values yield the value of $Q$, i.e., the amount of QCs of the state. On this purpose, we have presented the accessible methods and the necessary and sufficient resources to detect the amount of QCs of an unknown $2 \otimes d$ state for both the setups. In particular, in the NMR implementation, deeply entwined with the geometric perspective for QCs as we discussed in this work, the detection of $Q$  demands a number of local spin measurements smaller than tomography, with such a gain growing linearly with Bob's dimension. On the other hand, considering optical devices, we identified the circuits to be implemented for the measurements of $\{\langle O_i\rangle\}$:  swap or projective measurements over up to four copies of the state are required to detect QCs. \par
It is known that a single observable is sufficient to {\it detect} nonvanishing QCs \cite{laf2,mazi,cinesiwitness,dakic}. On the other hand, in general, we need a set of at least four independent measurements to {\it quantify} the amount of QCs for states of $2\otimes d$ systems.
The dimensionality of the Hilbert space of the system (in particular, the knowledge that subsystem $A$ is a qubit) is the only information we need {\it a priori} from the state, and this can be retrieved by a limited amount of supplemental resources \cite{scarani}.

The advantage of using $Q$ as a quantifier of QCs is even more evident when we are interested in dynamical contexts where the effort to reconstruct the state and therefore the correlations using standard tomographic approaches becomes even more demanding. Being a tight lower bound to geometric discord, $Q$ captures essentially the same features under decoherence dynamics, e.g. oscillations, sudden death and revivals. However attention must be payed in certain cases, e.g. sudden transitions in the dynamics of correlations, where instead $Q$ may manifest qualitatively different, typically smoother features.

Reaching tangible advances on the theoretical and experimental characterization of general QCs is relevant for quantum information theory, quantum foundations and the study of complex and many-body systems. We hope to have achieved a useful step forward in this scientific effort by providing new insights on QCs quantification. Nature is pervaded by evident and hidden manifestations of quantumness, and it seems legit to conjecture that some of its fundamental regulative processes are ruled and can be understood by electing QCs as primary tool of investigation.  For example, we could envisage that the QC dynamics in a bipartite system $S$ is  somewhat entwined with decoherence produced by the interaction of the system with an environment $E$.
Furthermore, even if it is still not conclusively known whether QCs in absence of entanglement can provide some actual appreciable help in developing better-than-classical protocols for any quantum information task, it is undoubtedly worthy to explore such a possibility. Experimentally sound recipes for measuring  QCs  for large dimensional systems, and serious engagement in studying their dynamics in the context of open quantum system, are pivotal to assess the role of QCs and their usefulness for the performance of such practical tasks in realistic conditions and, at the same time, to shed light on foundational questions of the broadest scientific prominence, such as the ultimate quantum picture of the measurement process.

\section*{Acknowledgements}
We thank the University of Nottingham for financial support through an Early Career Research and Knowledge Transfer Award, a Graduate School Travel Prize Award, and two EPSRC Research Development Fund projects (Grants ECRKTA/2011, TP/SEP11/10-11/181, RDF/PP/0312/09, and RDF/BtG/0612b/31). We warmly acknowledge discussions with Luigi Amico, Sougato Bose, Steve Clark, Borivoje Dakic, Radim Filip, Sevag Gharibian, Vittorio Giovannetti, Madalin Guta, Dieter Jaksch, Pawel Horodecki, Richard Jozsa, Chuan-Feng Li, Sabrina Maniscalco, Laura Mazzola, Ladislav Mista Jr., Tony Short, Diogo Soares-Pinto, Tommaso Tufarelli,  Vlatko Vedral, and Susy Virtuoso.  

\end{document}